\documentclass[journal,twoside,web]{ieeecolor}
\usepackage{generic}
\usepackage{amsmath,amssymb,amsfonts}
\usepackage{algorithmic}
\usepackage{graphicx}
\usepackage{textcomp}
\usepackage{booktabs}

\usepackage{float}
\usepackage{subfigure}
\usepackage{array}
\usepackage{multirow}
\usepackage{longtable}
\usepackage{rotating}
\usepackage{url}  
\newcommand{\RNum}[1]{\uppercase\expandafter{\romannumeral #1\relax}}

\def\BibTeX{{\rm B\kern-.05em{\sc i\kern-.025em b}\kern-.08em
    T\kern-.1667em\lower.7ex\hbox{E}\kern-.125emX}}
\begin{document}
%
\title{Dynamic radiomics: a new methodology to extract quantitative time-related features from tomographic images}
%
%
%
%

\author{Fengying Che, 
        Ruichuan Shi,
        Shuqin Li,
        Jian Wu,
        Haoran Li,         
        Weixing Chen,
        Hao Zhang,
        Hui Qu,  
        Zhi Li,
        and Xiaoyu Cui, \IEEEmembership{Member, IEEE} 
\thanks{F. Che, S. Li, H. Li, H. Qu and X. Cui are with College of Medicine and Biological Information Engineering, Northeastern University, Shenyang, Liaoning, China. X. Cui is with Key Laboratory of Intelligent Computing in Medical Image, Ministry of Education. J. Wu is with Key Laboratory of Data Analytics and Optimization for Smart Industry, Northeastern University, Shenyang, Liaoning, China. R. Shi, Z. Li are with Department of Medical Oncology,  the First Affiliated Hospital of China Medical University, Liaoning 110001, China. W. Chen is with Shenzhen College of Advanced Technology, University of the Chinese Academy of Sciences, Beijing 100049, China. H. Zhang is with Department of Breast Surgery, Liaoning Cancer Hospital and Institute, Cancer Hospital of China Medical University, Shenyang 110042, Liaoning, China.}
\thanks{F. Che and R. Shi have the same contribution to this paper.}
\thanks{X. Cui(E-mail: cuixy@bmie.neu.edu.cn) and Z. Li(Email: zli@cmu.edu.cn) are corresponding authors.}}

\maketitle

\begin{abstract}
The feature extraction methods of radiomics are mainly based on static tomographic images at a certain moment, while the occurrence and development of disease is a dynamic process that cannot be fully reflected by only static characteristics. This study proposes a new dynamic radiomics feature extraction workflow that uses time-dependent tomographic images of the same patient, focuses on the changes in image features over time, and then quantifies them as new dynamic features for diagnostic or prognostic evaluation. We first define the mathematical paradigm of dynamic radiomics and introduce three specific methods that can describe the transformation process of features over time. Three different clinical problems are used to validate the performance of the proposed dynamic feature with conventional 2D and 3D static features.Experimental results show that compared with static features, dynamic feature extraction can achieve higher robustness and accuracy for time-dependent tomographic images. We also found that the dynamic features that influence different clinical problems are also quite different.
\end{abstract}




%

\section{Introduction}\label{sec:introduction}

%
%
%
%
\IEEEPARstart{I}n the past decade, tomography imaging technologies (computed tomography (CT), magnetic resonance (MR), and positron emission tomography (PET)) have been widely used in clinical diagnosis, treatment planning and prognosis evaluation [1, 2]. Traditional experience-based diagnosis is easily influenced by subjective factors [3]. Thus, the quantitative analysis method based on radiomics has received widespread attention, through which a large number of high-throughput quantitative imaging features can be extracted and analyzed to carry out clinical decision making [4-6].

The core of radiomics is the extraction of high-dimensional feature data to quantitatively describe the properties of the region of interest (ROI). In general, there are two main categories of image features: handcrafted features and deep learning features. Handcrafted features mainly include first-order statistical features, texture features, shape features, and wavelet features. First-order statistics are derived from the histogram of voxel intensities [7], and statistical information such as the mean, median, skewness, and kurtosis can be calculated [8]. Texture features are global features that describe the structural properties between voxels. For example, the gray-level co-occurrence matrix (GLCM) [9], gray-level run length matrix (GLRLM) [10], gray-level size zone matrix (GLSZM) [11], and neighborhood gray-level different matrix (NGLDM) [12] have been proposed. Shape features have two representations: one is the contour feature, and the other is the region feature, which includes the area, perimeter, roundness, centroid, smallest rectangle containing the area of the mass, etc. Wavelet features refer to the characteristics of different frequency bands extracted from the wavelet decomposition of the image [3].

In addition, the use of image features extracted by deep learning for analysis and prediction is also common in the field of radiomics. In particular, convolutional neural networks (CNNs) have been extensively studied and used for deep learning feature extraction of different types of diseases [13-15]. Generally, deep learning features are extracted from a series of convolution filters of the convolution layer in the artificial neural network. Different convolution filters extract different features from the data and thus have different convolution layer outputs. The early diagnosis and prognosis of tumors are very important for the comprehensive treatment of patients. Ardila et al. developed a CNN cancer risk prediction model that uses a patient's current and prior low-dose computed tomography (LDCT) images to predict the risk of lung cancer [13]. In another study by Sajjad et al., the authors proposed a multigrade brain tumor classification system based on deep learning features to classify brain tumors into four different grades [14]. Moreover, Hao et al. used handcrafted features and deep learning features together from PET/CT, which were extracted from four different CNNs to quantify the tumor phenotype, and successfully stratified patients into a high-risk group and a low-risk group using a radiomics nomogram [15]. In recent years, various advanced structures of neural networks have been gradually applied in the field of radiomics. For instance, Xu and colleagues et al. used a dataset of 179 patients to develop a transfer learning CNN and recurrent neural network (RNN) model that integrates CT scans at multiple timepoints to predict response to lung cancer treatment, where RNN allowed the merging of several follow-up time points and was able to learn from samples with missed patient scans at a certain timepoint [16].

Acquiring more radiomics features is helpful to identify potential influencing factors that may be effective in predicting results. Until now, there has been a sustained effort to identify, define, and extract more radiomics features. Wu et al. [17] developed a sparse representation-based feature extraction method that exploits the statistical characteristics of the lesion area, which could be used for the outcome prediction of higher-grade gliomas in the future [18]. Based on graph theory, the features extracted by Zhou et al. [19] can specifically represent PET image characteristics. Moreover, the extraction of tomography image features can also be extended to the three-dimensional domain [20]. For some specific applications, researchers have found that 3D features can achieve better performance than 2D features [21]. Meanwhile, 2D and 3D features are often used in combination and have shown a better performance than that from 2D or 3D alone [22]. In the field of deep learning features, Dai et al. proposed TransMed, a new transformer-based multimodal medical image classification method, which combines the advantages of CNN and transformer to efficiently extract low-level features of images and establish long-range dependencies between modalities [23]. These features can potentially be extracted from individual habitats, thereby yielding thousands of data elements with which to describe each volume of interest, with many volumes of interest available for each patient [24].

Regardless of which feature extraction technology is adopted, the data processed by the existing methods are mainly based on static tomographic images at a certain moment. However, the occurrence and development of tumors is a dynamic process that cannot be fully reflected by only static characteristics. In some clinical applications, time-series images or images from multiple periods are required for diagnosis or prognosis. On the one hand, current research analysis shows that pharmacokinetics is suitable for the clinical analysis and prediction of a variety of cancers [25-27]; thus, dynamic contrast-enhanced magnetic resonance imaging (DCE-MRI) is commonly used to analyze the metabolic processes of the tumor in patients suspected of having breast cancer or prostate cancer [28,29]. On the other hand, the size and shape of the tumor can be detected by tomography directly; thus, imaging changes at different treatment stages are also used to assess the efficacy of radiotherapy or neoadjuvant chemotherapy in cancer patients [30]. Recently, Carvalho et al. proposed “delta radiomics”, which can express the rate of change of a radiomics feature over time [31]. This approach can provide additional information to identify, quantify, and potentially predict treatment-induced changes during treatment and has been shown to have potential in evaluating the efficacy of colorectal [30], liver [32], pancreatic [33] and lung cancers [34]. However, these time-related diagnostic procedures are still in the initial stage, so a rigorous model is still lacking and the procedures have not been well established in the existing radiomics techniques.

In this study, we propose a novel dynamic radiomics feature extraction workflow that can take advantage of time-related tomography images of the same patient by paying attention to the changes in image features over time and then quantifying them into new dynamic features for diagnosis and prognosis assessment. We did not use deep learning features here mainly for the following two reasons [16, 35]: (1) deep learning features are often not interpretable, while handcrafted features have been widely recognized by clinicians; (2) the extraction of depth features requires a large number of samples, and it is difficult to obtain good results for a small sample set. First, we introduce the basic framework of static radiomics feature extraction and then propose the definition of dynamic radiomics features. In addition, we introduce three specific methods that can describe the transformation process of features over time. A total of 241 samples from three different clinical problems are used to validate the performance of the proposed feature extraction method.

The first problem is the prediction of axillary lymph node metastasis (ALNM) in breast cancer, which is one of the most important elements that can affect prognosis [36]. Many efforts have been made to develop a prediction model based on static radiomics features, such as using T2-weighted MRI images [37] or the first phase of T1-DCE images [38]; however, it is difficult for the predictive accuracy to meet clinical requirements. In this paper, we exploit the improved prediction performance based on the complete time-series images of DCE-MRI by using dynamic radiomics features.

The second problem is the prediction of gene mutation status as a noninvasive strategy. Previous studies [39] have reported that resistance to the anti-EGFR antibody cetuximab was caused by a clone with a pre-existing KRAS mutation and/or a mutation generated during treatment. Both mechanisms of resistance may occur concomitantly. Radiomics features can be used for detection multiple times during the whole treatment process, especially at each key treatment node. In this way, the changes in gene mutation status can be tracked at any time to track the changes in tumor heterogeneity and guide clinical decisions. In this study, we aimed to construct a model based on radiomic features obtained from multiple phases to improve the noninvasive assessment of RAS and BRAF mutations in patients with colorectal cancer liver metastasis (CRLM) prior to any treatment.

The third problem is the prediction of neoadjuvant chemotherapy. Clinically, it is of great significance to predict the efficacy of neoadjuvant chemotherapy, and the ability to identify nonresponders early may allow the selection of patients who may benefit from a therapy change. Current studies [40] have shown that some pharmacokinetic parameters are significantly related to the efficacy of neoadjuvant chemotherapy. In addition, its curative effect can be well predicted through traditional static radiomics methods [41]. In this study, we used public data [42] to verify the dynamic radiomics method.

The remainder of the paper is organized as follows. Section II provides some background on the static radiomics features. Section III presents the proposed time-related dynamic radiomics feature extraction method, which contains discrete time-related feature extraction, integrated time-related feature extraction and parameter fitting feature extraction. Section IV reports the experimental results of the proposed method for differentiating sentinel lymph node (SLN) metastasis of breast cancer, for estimating liver cancer gene mutations and for predicting the effects of neoadjuvant chemotherapy. Section V concludes the proposed method.


 

\section{STATIC RADIOMICS FEATURES}

At present, radiomics is mostly based on medical tomographic images at a particular moment to extract the characteristics of the ROI. Therefore, in this study, we describe the existing radiomics technology as static radiomics, and its corresponding radiomics features can be specifically expressed as:

\begin{equation}
F=\{\psi(x(t))\in \mathbb{R} | x(t)\in \mathbb{R}^{m\times n\times p},t\in \mathbb{R}^+ \}
\label{eq.1}
\end{equation}

where $x(t)$ represents the area of interest extracted from the tomographic images collected at time $t$, and the function $\psi$  denotes the feature extraction method (including first-order statistical features, texture features, shape features, complex features, etc. The details are shown in Appendix I.), and $p$ represents the number of layers in the tomographic images. When $p$ is equal to one, 2D features are extracted; when $p$ is greater than one, 3D features are extracted. Suppose the number of extracted features is $q$. The feature extraction process of traditional static radiomics is essentially a mapping process from $m\times n\times p$ dimensional tensor to $q$ dimensional space.

\section{DYNAMIC RADIOMICS FEATURES}

In this study, we propose the concept of dynamic radiomics for the first time. Its purpose is to construct new time-related features that can describe the change rule by taking advantage of the static characteristic changes at different time points, which can be expressed as:

\begin{equation}
\phi(\psi(x(t_1)), \psi(x(t_2)), \cdots, \psi(x(t_k)))
\label{eq.2}
\end{equation}

where $\phi(\cdot)$ denotes the transformation from $\mathbb{R}^k$ to $\mathbb{R}^d$; here, $d$ is the number of dynamic features that can be extracted. According to the number of time points collected and the method of feature extraction, we propose three kinds of methods for calculating time dimension features, namely, integrated features, discrete features and parameter features. The details are shown in Appendix II.

\subsection{Integrated Feature}

According to the statistical analysis model, this method uses the characteristics of each time point to build a model that can describe the overall change rule of the characteristics as the characteristics of the time dimension. For example, $\phi(x_1,x_2,…,x_k)$ can denote the statistical function of a sample, such as the mean, variance, or coefficient of variation. 

When

\begin{equation}
\phi(x_1,x_2, \cdots,x_k)=\frac{1}{k} \sum_{i=1}^k x_i
\label{eq.3}
\end{equation}

After the transformation of the function $\phi(\cdot)$, the dynamic features can be expressed as the average of the features at all times:

\begin{equation}
\phi(\psi(x(t_1)), \psi(x(t_2)), \cdots, \psi(x(t_k)))=\frac{1}{k} \sum_{i=1}^k \psi(x(t_i))
\label{eq.4}
\end{equation}

When

\begin{equation}
\phi(x_1,x_2, \cdots,x_k)=\frac{1}{k} \sum_{i=1}^k |x_i-\frac{1}{k} \sum_{i=1}^k x_i|
\label{eq.5}
\end{equation}

the dynamic features can be express as:

\begin{equation}
\begin{split}
&\phi(\psi(x(t_1)), \psi(x(t_2)), \cdots, \psi(x(t_k))) \\
&=\frac{1}{k} \sum_{i=1}^k |\psi(x(t_i))-\frac{1}{k} \sum_{i=1}^k \psi(x(t_i))|
\label{eq.6}
\end{split}
\end{equation}

Here, $\phi(\cdot)$ can also be other types of statistical functions. In the process of tumor diagnosis or treatment, different imaging features change constantly with drug metabolism or treatment progress. Integrated features are used to quantitatively describe the overall law of these changes. 

\subsection{Discrete Feature}

If $\phi(\cdot)$ denotes the transformation from $\mathbb{R}^k$ to $\mathbb{R}^d$, where $d=k(k-1)/2$, then the transformation can be represented by the following operator: $M=(m_{ij})_{k\times k}$ , which is a matrix of $k\times k$ dimensions; $g(\cdot,\cdot) \mathbb{R}^2 \to \mathbb{R}^1$ is a function, where

\begin{equation}
m_{ij}=g(\psi(x(t_i)), \psi(x(t_j))), 1 \le i, j \le k
\label{eq.7}
\end{equation}

Define $P$ as an operation that takes the characteristic elements of the upper triangular region that do not contain diagonals of $M$ and then straightens them; thus, we have $PM\in \mathbb{R}^{k(k-1)/2}$, which can be used as dynamic features. For example, when $g(x,y)=|x-y|/y$, the dynamic features can be expressed as:

\begin{equation}
\begin{split}
\phi=&(\frac{|\psi(x(t_1))-\psi(x(t_2))|}{\psi(x(t_2))},\frac{|\psi(x(t_1))-\psi(x(t_3))|}{\psi(x(t_3))},\\
&\cdots,\frac{|\psi(x(t_{k-1}))-\psi(x(t_k))|}{\psi(x(t_k))})
\label{eq.8}
\end{split}
\end{equation}

This method uses the interaction effect of the characteristics of two points in time, constructs the function to describe the characteristic change rule of two points, and runs the two points through the whole time point. Thus, the discrete feature can compensate for the deficiency of the integrated feature extraction method in detail feature extraction and quantify the characteristic change between each time point.For example, delta radiomics uses variance to evaluate the rate of change (when t=2), which belongs to the category of discrete features.

\subsection{Parameter Feature}
For the feature data corresponding to $k$ moments, the corresponding data of $k$ groups can generally be obtained as:

\begin{equation}
\{(t_i,\psi(x(t_i))), i=1,2,\cdots,k\}
\label{eq.9}
\end{equation}

To describe the variation rule of extracted features with time at different moments, the common parameter model in statistics is used, and corresponding parameters can also be used as extracted features. For example, when using the following parametric model:

\begin{equation}
\psi(x(t_i))=m(t_i,\theta)+\varepsilon_i, i=1,2,\cdots,k
\label{eq.10}
\end{equation}

Where $\theta \in \mathbb{R}^d$ represents unknown parameters of the model, $m(\cdot)$ is a given function, $\varepsilon_i$ represents the random error, and $E\varepsilon_i=0$, Based on the least square estimation model, the datasets $\{(t_i, \psi(x(t_i))),i=1,2,\dots,k\}$ can be solved by the following optimization problem:

\begin{equation}
\hat{\theta}=\mathop{\arg\min}_{\theta \in \Theta} \ \ \ \sum_{i=1}^{k}(\psi(x(t_i))-m(t_i,\theta))^2
\label{eq.11}
\end{equation}

Here, the dynamic radiomics features are the parameter $\hat{\theta}$. Below are some examples of models and their corresponding dynamic radiomics features:

\begin{equation}
\left\{
             \begin{array}{lr}
             m(t_i,\theta)=\frac{A\cdot(1-e^{-\alpha \cdot t})^q\cdot e^{-\beta \cdot t}\cdot(1+e^{-\gamma \cdot t})}{2}, \theta=(A,\alpha,q,\beta,\gamma)^T &  \\
             \\
             m(t_i,\theta)=\sum_{i=1}^{7}a_i\cdot t^i,\theta=(a_1,a_2,\cdots,a_7)^T\\
             \\
             m(t_i,\theta)=\frac{(P_2+(P_5\cdot t))}{(1+e^{-P_4(t-P_3)})}+P_1,\theta=(P_1,P_2,\cdots,P_5)^T, &  
             \end{array}
\right.
\end{equation}

The parameter feature refers to the idea of pharmacokinetics, the characteristic values at each moment are fitted with a specific curve, and the fitted parameters are used as the results. The information content of this method is between the integrated characteristics and the discrete characteristics. It can describe both global change and local change, which is suitable for data analysis at many time points.

\section{EXPERIMENT AND ANALYSIS OF RESULTS}

\begin{table*}[htbp]
\centering
\caption{\label{tab:test}Equipment Parameters of Different Cohorts. Cohort 1 is the DCE-MRI data of breast cancer patients from the First Affiliated Hospital of China Medical University. Cohort 2 is the DCE-MRI data of breast cancer patients from Shengjing Hospital of China Medical University. Cohort 3 is the CT data of patients with liver metastases from bowel cancer. Cohort 4 is the DCE-MRI data of patients treated with neoadjuvant chemotherapy from a public database.}
\setlength{\tabcolsep}{14mm}
\begin{tabular}{ll}
\toprule
Cohort 1 & Cohort 2 \\
\midrule
Manufacturers: Siemens 3.0 T MRI & Manufacturers: Philips 3.0 T MRI \\
TR: $4.46\sim 7.80 ms$ & TR: $4.1 ms$ \\
TE: $1.54\sim 4.20 ms$ & TE: $2.1 ms$ \\
No interval scanning & No interval scanning \\
Slice thickness: $2.0 mm$ & Slice thickness: $2.0 mm$ \\
Stages in the scan: 8 & Stages in the scan: 8 \\
Interval between stages: 1 min & Interval between stages: 1 min \\
\toprule
Cohort 3 & Cohort 4\\
\midrule
Manufacturers: Toshiba, GE, Phillips and Siemens & Manufacturers: Siemens 3.0 T MRI \\
Tube voltage: 120 kVp (range $100-140 kVp$) & TR: 6.2 ms \\
Slice thickness: 2.0 mm & TE: 2.9 ms \\
Matrix: $512\times 512$ & FOV: $30\sim 34 cm$ \\
Tube current: 333 mA (range 100–752 mA) & Slice thickness: 1.4 mm \\
Exposure time: 751 ms (range 500–1782 ms) & In-plane matrix size: $320\times 320$ \\
\bottomrule
\end{tabular}
\\TE: echo time, TR: pulse repetition time, FOV: field of view
\end{table*}

\begin{table*}[htbp]
\centering
\caption{\label{tab:test}The Parameters of Cohort 1.}
\setlength{\tabcolsep}{7mm}
\begin{tabular}{llll}
\toprule
Characteristics & Metastasis (n = 30) & Nonmetastasis (n = 27) & P value \\
\midrule
Histological grade &  &  & \\
\hspace{3cm}1& 1&  0 & \\
\hspace{3cm}2& 25& 26 & \\
\hspace{3cm}3& 4& 1 & \\
Stage &  &  & $1.212*e-8$\\
\hspace{3cm}\uppercase\expandafter{\romannumeral1}+\uppercase\expandafter{\romannumeral2}& 7& 26 & \\
\hspace{3cm}\uppercase\expandafter{\romannumeral3}+\uppercase\expandafter{\romannumeral4}& 23& 1 & \\
Molecular subtype & & & \\
\hspace{3cm}Luminal A & $4(13.33\%)$ & $8(29.63\%)$ & \\
\hspace{3cm}Luminal B & $18(60\%)$ & $15(55.56\%)$ & \\
\hspace{3cm}HER2-like & $1(3.33\%)$ & $1(3.7\%)$ & \\
\hspace{3cm}Triple-negative & $7(23.33\%)$ & $3(11.11\%)$ & \\
Age & & & 0.1135\\
\hspace{3cm}$\le$ median & $12(40\%)$ & $17(62.96\%)$	\\
\hspace{3cm}$>$ median	& $18(60\%)$ & $10(37.04\%)$ \\
Tumor size, cm & & & 0.001059\\
\hspace{3cm}$\le$2 & 3 & 14 \\
\hspace{3cm}$>$2 & 27 & 13 \\
\bottomrule
\end{tabular}
\\The exact P values and Kruskal-Wallis test were used to determine whether the clinicopathological variables differed significantly between the metastasis and nonmetastasis sets..
\end{table*}

\begin{table*}[htbp]
\centering
\caption{\label{tab:test}The Parameters of Cohort 2.}
\setlength{\tabcolsep}{8mm}
\begin{tabular}{llll}
\toprule
Characteristics & Metastasis (n = 34) & Nonmetastasis (n = 33) & P value \\
\midrule
Stage &  &  & $0.0003732$\\
\hspace{3cm}\uppercase\expandafter{\romannumeral1}+\uppercase\expandafter{\romannumeral2}& 23& 33 & \\
\hspace{3cm}\uppercase\expandafter{\romannumeral3}+\uppercase\expandafter{\romannumeral4}& 11& 0 & \\
Molecular subtype & & & \\
\hspace{3cm}A & 3 & 7 & \\
\hspace{3cm}B & 24 & 19 & \\
\hspace{3cm}Her2+ & 3 & 2 & \\
\hspace{3cm}TNBC & 4 & 3 & \\
\hspace{3cm}unknown & 0 & 2 & \\
Age & & & 0.8086\\
\hspace{3cm}$\le$ median & $18(52.94\%)$ & $16(48.48\%)$	\\
\hspace{3cm}$>$ median	& $16(47.06\%)$ & $17(51.52\%)$ \\
Tumor size, cm & & & 0.02803\\
\hspace{3cm}$\le$2 & 11 & 20 \\
\hspace{3cm}$>$2 & 23 & 13 \\
\bottomrule
\end{tabular}
\\The exact P values and Kruskal-Wallis test were used to determine whether the clinicopathological variables differed significantly between the metastasis and nonmetastasis sets.
\end{table*}

\begin{table*}[htbp]
\centering
\caption{\label{tab:test}The Parameters of Cohort 4.}
\setlength{\tabcolsep}{9mm}
\begin{tabular}{llll}
\toprule
Characteristics & Mutant type (n = 64) & Wild type (n = 43)  & P value \\
\midrule

Microsatellite &  &  & 0.4205\\
\hspace{3cm}no & $39(60.94\%)$ & $35(81.4\%)$\\
\hspace{3cm}yes & $25(39.06\%)$ & $8(18.6\%)$\\

Extrahepatic Meta &  &  & 0.006\\
\hspace{3cm}no & $38(59.38\%)$ & $37(86.05\%)$\\
\hspace{3cm}yes & $26(40.62\%)$ & $6(13.95\%)$\\

Sex & & & 0.202\\
\hspace{3cm}female & $30(46.88\%)$ & $14(32.56\%)$ \\
\hspace{3cm}male & $34(53.12\%)$ & $29(67.44\%)$ \\

Age & & & 1\\
\hspace{3cm}$\le$60 & $33(51.56\%)$ & $23(53.49\%)$ \\
\hspace{3cm}$>$60 & $31(48.44\%)$ & $20(46.51\%)$ \\

Tumor site & & & \\
\hspace{3cm}left &      $19(29.69\%)$ & $11(26.19\%)$ \\
\hspace{3cm}rectum &    $26(40.62\%)$ & $19(45.24\%)$ \\
\hspace{3cm}right &     $14(21.88\%)$ & $6(14.29\%)$ \\
\hspace{3cm}unknown &   $5(7.81\%)$ & $6(14.29\%)$ \\

\bottomrule
\end{tabular}
\\“Extrahepatic Meta” refers to the presence of metastatic lesions other than the liver and regional lymph nodes. “Microsatellite” refers to a single large lesion surrounded by multiple small lesions. The exact P values and Kruskal-Wallis test were used to determine whether the clinicopathological variables differed significantly between the metastasis and nonmetastasis sets.
\end{table*}

\begin{table*}[htbp]
\centering
\caption{\label{tab:test}Model built using data from breast cancer patients, with the prediction accuracy based on various dimensions of FNN, SVM and LDA. The RCR, RACA, SD, DC and Ploy’s formulas are in Appendix II, where 2D* represents 2D dynamic features and 3D* represents 3D dynamic features. The static feature refers to separately using 2D and 3D features for modeling and analysis, where 2D Multi and 3D Multi refer to feature set analysis at multiple time points.}
\setlength{\tabcolsep}{9mm}
\begin{tabular}{lllll}
\toprule
 & & FNN & SVM  & LDA \\
\midrule
Discrete Feature & $3D^*\_RCR$ & $1.000$ & $0.952$ & $0.833$ \\

        &   $3D^*\_RACR$ &  $0.952$ &   $0.857$ & $0.929$\\
        &   $2D^*\_RCR$ &   $0.881$ &   $0.857$ & $0.857$\\
        &   $2D^*\_RACR$ &  $0.857$ &   $0.881$ & $0.976$\\

\midrule
Integrated Feature &  $3D^*\_SD$ & $0.667$ & $0.592$ & $0.619$\\
	             &  $3D^*\_DC$ & $0.714$ & $0.643$ & $0.619$\\
	             &  $2D^*\_SD$ & $0.810$ & $0.738$ & $0.786$\\
	             &  $2D^*\_DC$ & $0.786$ & $0.738$ & $0.738$\\

\midrule	             
Parameter Feature & $3D\_Ploy$ & $0.425$ & $0.450$ & $0.625$\\
	              & $2D\_Ploy$ & $0.300$ & $0.275$ & $0.325$\\

\midrule
Static Feature &  $3D\_Multi$ & $0.857$ & $0.595$	& $0.810$\\
	         &  $2D\_Multi$ & $0.833$ & $0.881$	& $0.833$\\
	         &  $3D$        & $0.600$ & $0.628$	& $0.684$\\
	         &  $2D$        & $0.500$ & $0.356$	& $0.453$\\

\bottomrule
\end{tabular}
\end{table*}

\subsection{Data acquisition}
This study used three types of data according to the three different clinical problems, as shown in Table \RNum{1}. For the first problem, the data of breast cancer patients came from two hospitals: 57 patients were from the First Affiliated Hospital of China Medical University, and 67 patients were from Shengjing Hospital of China Medical University; the resolutions of the images from the two hospitals were 384*384 and 560*560, respectively. The data were divided into two groups (64 SLN-positive cases and 60 SLN-negative cases). For the second problem, the data of 107 patients (63 men and 44 women, median age of 60 years, age ranging between 35 and 78 years) from the First Affiliated Hospital of China Medical University were collected for the CRLM study. Based on the status of the RAS and BRAF genes, the patients were classified into two groups: the mutant group and the wild-type group. Patients with any mutations in the RAS or BRAF gene were classified into the mutant group (N = 64), while others were classified into the wild-type group (N = 43). For the third problem, an open data set was used, and data were gathered prior to the start of treatment (V1) and after the first cycle of treatment (V2) from 10 patients: 3 pathological complete responders (PCRs) and 7 non-PCRs (\url{https://wiki.cancerimagingarchive.net/display/Public/QIN+Breast+DCE-MRI}). The resolution of the images was 320*320. The goal of the challenge was to evaluate variations in the DCE-MRI assessment of breast cancer response to neoadjuvant chemotherapy. The scanning parameters are listed in Tables \RNum{2}-\RNum{4}. A pathological diagnosis was made and confirmed by authoritative doctors. This study was approved by the Ethics Committee of the First Affiliated Hospital of China Medical University and the Ethics Committee of the Shengjing Hospital of China Medical University.

\subsection{Work pipeline}
The image biomarker standardization initiative (IBSI) [43] was regarded as a reference and taken into consideration in our data processing, image feature, and biomarker selection procedure. The experimental processing flow of this study is summarized as follows: (1) acquisition of image data; (2) tumor area calibration; (3) tumor area segmentation; (4) multiperiod static feature extraction and quantification; (5) dynamic feature generation and dimensionality reduction; and (6) dynamic classification model establishment and prediction. The following is a corresponding introduction to the process and challenges. First, we obtained multicenter experimental data and used the ROI outlined by the clinician to segment the tumor region. Furthermore, static features such as first-order statistical features, texture features, and shape features were extracted from the segmented ROI in the image. Then, we used the dynamic algorithm we designed to transform the static features into dynamic features and used the least absolute shrinkage and selection operator (LASSO) method for feature selection. Finally, the features after dimensionality reduction were used for modeling, and in this process, we used the Wu Kong platform (https://www.omicsolution.com/wkomics/main/) for relative analysis. And methods such as receiver operating characteristic (ROC) curves were used for model evaluation. Among them, we used the holdout method to divide the data into a training set and a test set at a ratio of 2:1. 

\subsection{Prediction of ALNM in breast cancer and comparison of the predictive effect on data from different sources}

In this part of the work, we first conducted experiments using data from breast cancer patientsTo fully prove the validity of dynamic features, the prediction model was established by using three classifiers (feedforward neural network (FNN), support vector machine (SVM) and linear discriminant analysis (LDA)) with the best classification effect. We compared and analyzed three types of dynamic models and one static model in various dimensions. In all model building processes, we used two-thirds of the data as the training set and one-third of the data as the test set. The results are shown in Table \RNum{5}.

It can be seen from the information in Table \RNum{5} that compared to the static model, the discrete feature prediction accuracy rate in the dynamic model is greatly improved. Among them, for the $3D^*\_RCR$ algorithm, the FNN model and the SVM model have the best prediction results. For the $3D^*\_RACR$ algorithm, the FNN model and the LDA model have the best results. In contrast, the $2D^*\_RCR$ algorithm did not perform very well under the three classifier models. The $2D^*\_RACR$ algorithm performed better under the LDA model. For the static models, the analysis and prediction results using single-period 2D and 3D features are relatively general, and the analysis and prediction results using multiperiod 2D and 3D features are better but still worse than the prediction results of the discrete feature in the dynamic model. However, in the dynamic model, the integrated feature and parameter features have general results in breast cancer prediction. In general, the models established by dynamic 2D and 3D radiomics methods are significantly better than traditional static 2D and 3D radiomics models. At the same time, dynamic 3D prediction models have higher accuracy than dynamic 2D models. This finding suggests that dynamic 3D features based on medical images are more suitable than dynamic 2D features to describe the heterogeneity of the lesion area and can more fully describe the metabolic changes of the lesion. For the discrete feature in the dynamic model, which had the best results, we used the ROC curve to evaluate the performance of the models built by these two algorithms in different dimensions. The ROC curve results are shown in Fig. 1.

\bgroup
\begin{figure*}[htbp]
\centering \makeatletter\includegraphics[width=7.00in, keepaspectratio]{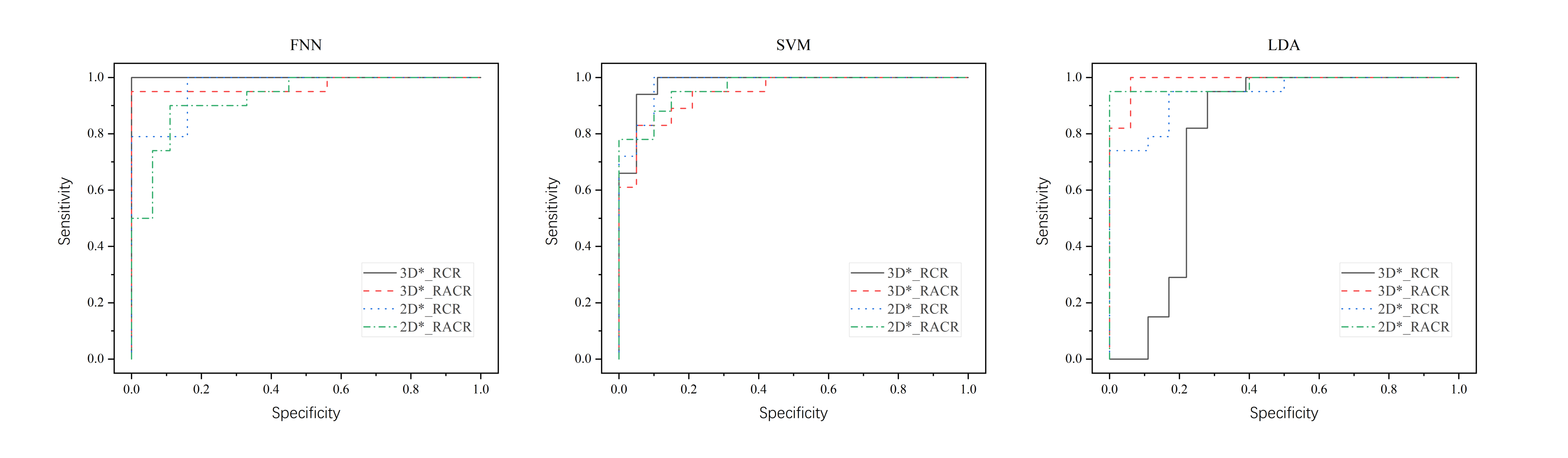}{}
\makeatother 
\caption{{The ROC curves of the FNN, SVM and LDA models for each dimension.}}
\label{Network}
\end{figure*}
\egroup

\bgroup
\begin{figure*}[htbp]
\centering \makeatletter\includegraphics[width=5.50in, keepaspectratio]{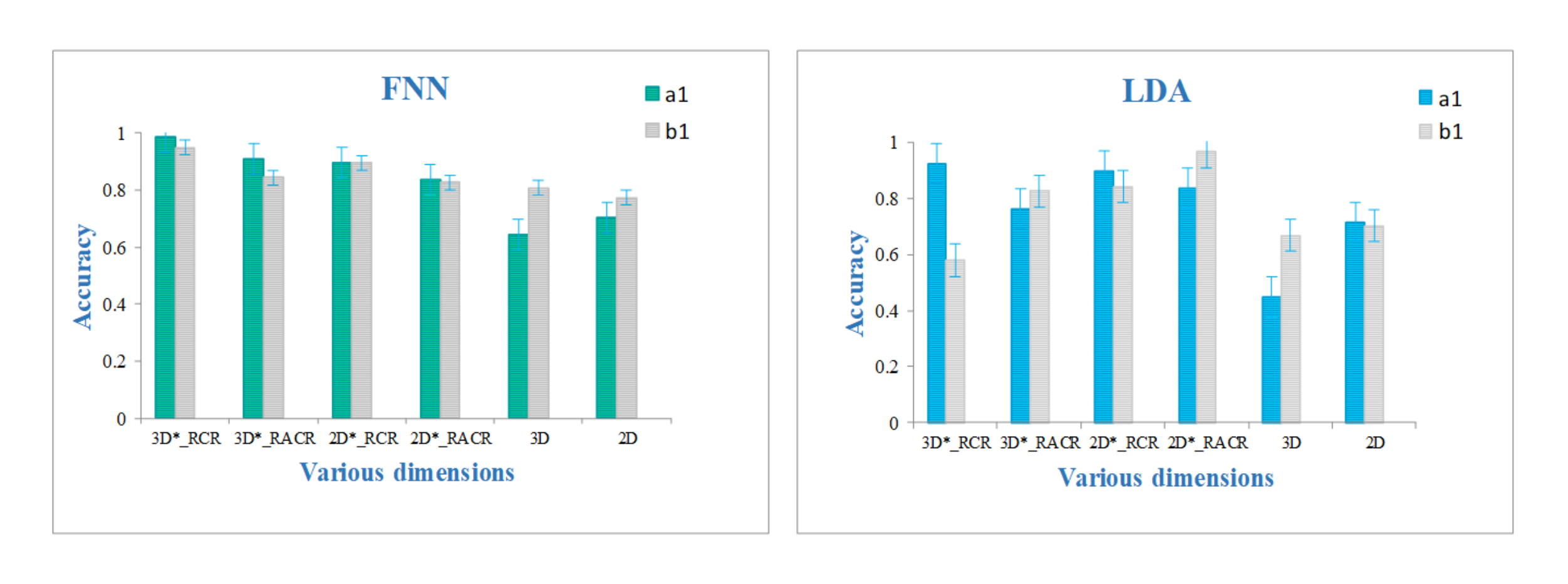}{}
\makeatother 
\caption{{The accuracy of FNN and LDA models for each dimension. a1 is the accuracy of the model using the data from the First Affiliated Hospital of China Medical University as the training set and the data from Shengjing Hospital as the test set. b1 is the accuracy rate of the model established by interchanging the training set and the test set.}}
\label{Network}
\end{figure*}
\egroup

Since the data of breast cancer patients were from a multicenter data set, there were some differences in the data obtained. These differences are mainly reflected in the difference in the resolution of DCE-MRI, which also directly affects the dynamic analysis of radiomics features. Therefore, in this section, we compare and analyze the data of the two hospitals to evaluate the generalization of the new algorithm, which is named the discrete feature, on data from different sources. This part also uses two models of FNN and LDA. The data from the two hospitals are separately used as the training and test sets for experiments. The results are shown in Fig. 2.

\bgroup
\begin{figure*}[htbp]
\centering \makeatletter\includegraphics[width=6in, keepaspectratio]{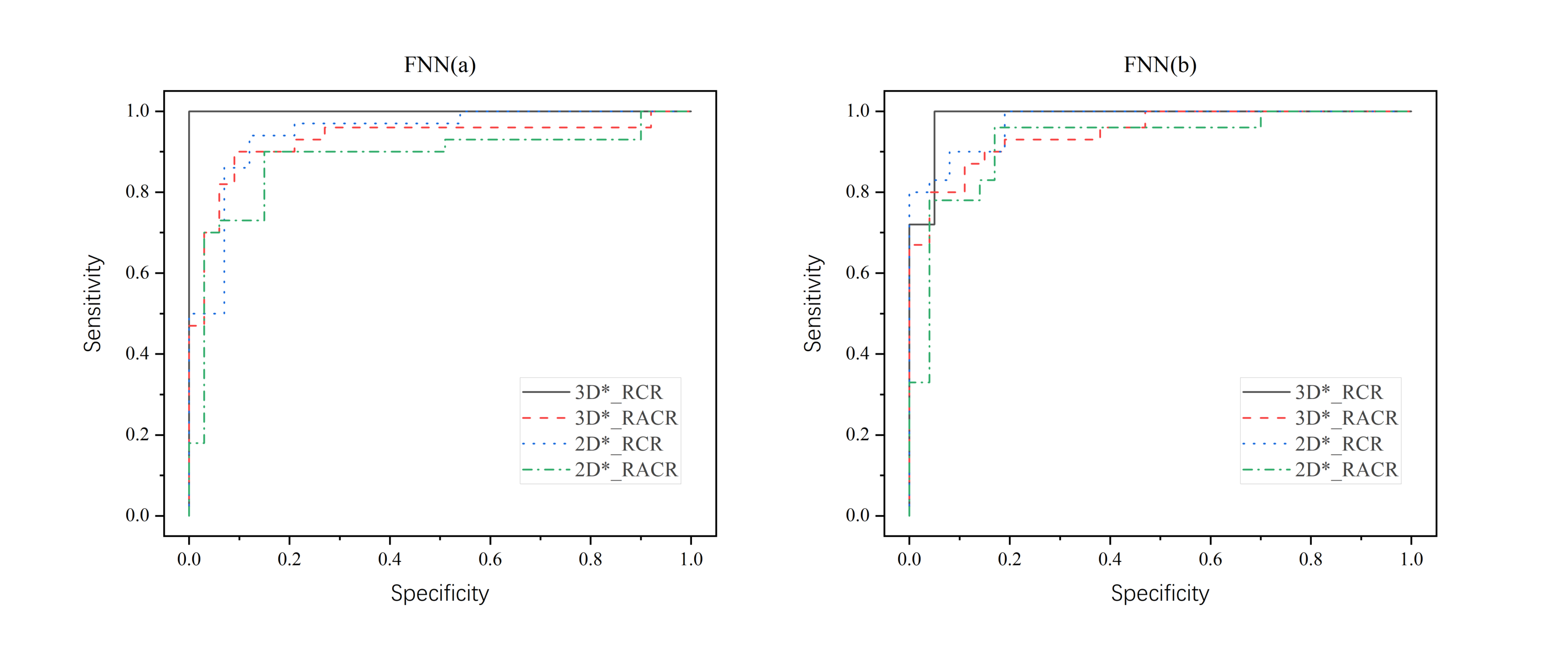}{}
\makeatother 
\caption{{FNN(a) is the ROC curve of the FNN model for each dimension, which uses the data from the First Affiliated Hospital of China Medical University as the training set and the data from Shengjing Hospital as the test set. FNN(b) is the ROC curve of the FNN model for each dimension, which uses the data from Shengjing Hospital as the training set and the data from the First Affiliated Hospital of China Medical University as the test set.}}
\label{Network}
\end{figure*}
\egroup

The analysis and comparative experiments of different data sources show that for the FNN model, the model established by dynamic 2D and 3D radiomics algorithms can eliminate the influence of different data sources to a certain extent. For data from two data sources, regardless of which set of data is used to train the model and which is used to test the model, the prediction results of the model are almost the same. The algorithm with the best performance among the four dynamic algorithms is $3D^*\_RCR$. In contrast, the accuracy and stability of traditional static 2D and 3D models are not very good. In the LDA model, among the four dynamic algorithms, $3D^*\_RACR$, $2D^*\_RCR$ and $2D^*\_RACR$ have better stability. The stability of 2D features in the static method is better, but the accuracy is still not as good as that of the dynamic algorithm. In general, models built using traditional static 2D and 3D features are not suitable for predicting data from different data sources. In other words, static 2D and 3D radiomics methods are more suitable for predicting image data with no difference in resolution, while dynamic 2D and 3D radiomics methods are more suitable for data with differences. For the FNN model, which had the best results, we used the ROC curve to evaluate its prediction performance. The ROC curve results are shown in Fig. 3.

\subsection{Prediction of gene mutation status in patients with liver metastases from bowel cancer}

In addition, we also used data from patients with liver metastases from bowel cancer to experiment with this dynamic radiomics method. Radiomic features were extracted from the precontrast phase (PP), arterial phase (AP), portal venous phase (DP) and delay phase. The results of the experiment are shown in Table \RNum{6}.

\begin{table*}[htbp]
\centering
\caption{\label{tab:test}Model built using data from patients with liver metastases from bowel cancer, with the prediction accuracy based on various dimensions of FNN, SVM and LDA. The RCR, RACA, SD, DC and Ploy’s formulas are in Appendix II, where 2D* represents 2D dynamic features and 3D* represents 3D dynamic features. The static feature refers to separately using 2D and 3D features for modeling and analysis, where 2D Multi and 3D Multi refer to feature set analysis at multiple time points.}
\setlength{\tabcolsep}{9mm}
\begin{tabular}{lllll}
\toprule
 & & FNN & SVM  & LDA \\
\midrule
Discrete Feature & $3D^*\_RCR$	& $0.556$	& $0.500$	& $0.556$ \\
                   & $3D^*\_RACR$	& $0.528$	& $0.583$	& $0.556$ \\
                   & $2D^*\_RCR$	& $0.611$	& $0.528$	& $0.722$ \\
                   & $2D^*\_RACR$	& $0.694$	& $0.667$	& $0.778$ \\

\midrule
Integrated Feature & $3D^*\_SD$ & $0.639$	& $0.556$ & $0.611$ \\
                 & $3D^*\_DC$ & $0.806$	& $0.806$ & $0.833$ \\
                 & $2D^*\_SD$ & $0.500$	& $0.500$ & $0.700$ \\
                 & $2D^*\_DC$ & $0.528$	& $0.528$ & $0.806$ \\

\midrule	             
Parameter Feature & $3D\_Ploy$ & $0.619$ & $0.619$ & $0.690$ \\
                  & $2D\_Ploy$ & $0.548$ & $0.500$ & $0.476$ \\

\midrule
Static Feature & $3D\_Multi$ & $0.528$ & $0.556$ & $0.611$ \\
             & $2D\_Multi$ & $0.639$ & $0.500$ & $0.528$ \\
             & $3D$	  & $0.681$ & $0.639$ & $0.651$ \\
             & $2D$	  & $0.528$ & $0.611$ & $0.527$ \\
\bottomrule
\end{tabular}
\end{table*}

In this part of the experimental results, the $3D^*\_DC$ algorithm in the dynamic method named integrated feature performs better. In contrast, discrete feature, which performed best in predicting breast cancer, did not perform well in this problem. This means that for different clinical diagnoses, different characteristics will show a higher correlation, rather than a certain type of characteristic being applicable to all clinical diagnoses. The predictive effect of traditional static 2D and 3D methods in this part of the clinical data is not very good.

\bgroup
\begin{figure}[htbp]
\centering \makeatletter\includegraphics[width=3.5in, keepaspectratio]{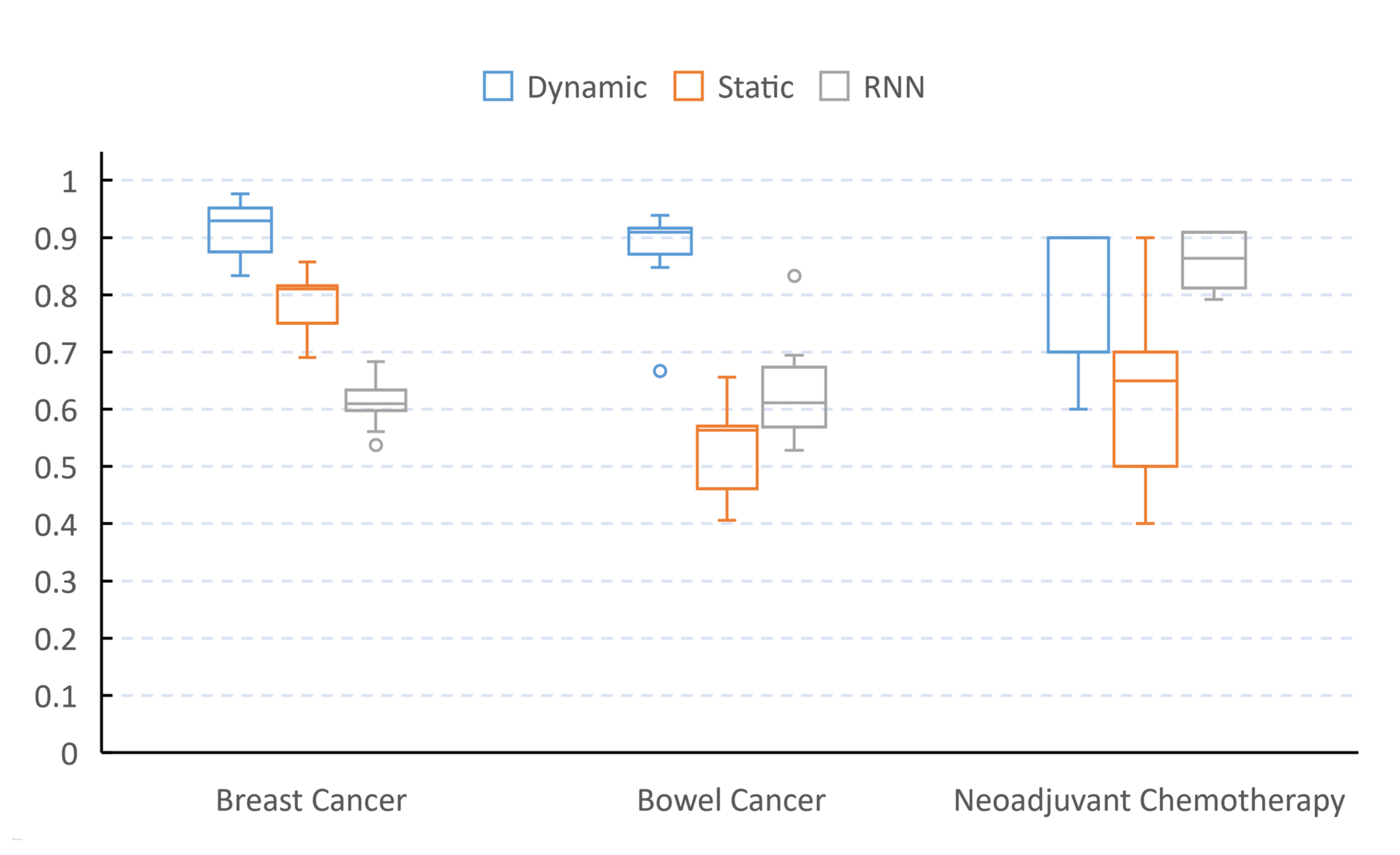}{}
\makeatother 
\caption{{The accuracy of dynamic model, static model and RNN model. Among them, the dynamic model and the static model respectively use FNN.}}
\label{Network}
\end{figure}
\egroup

\bgroup
\begin{figure*}[htbp]
\centering \makeatletter\includegraphics[width=7.00in, keepaspectratio]{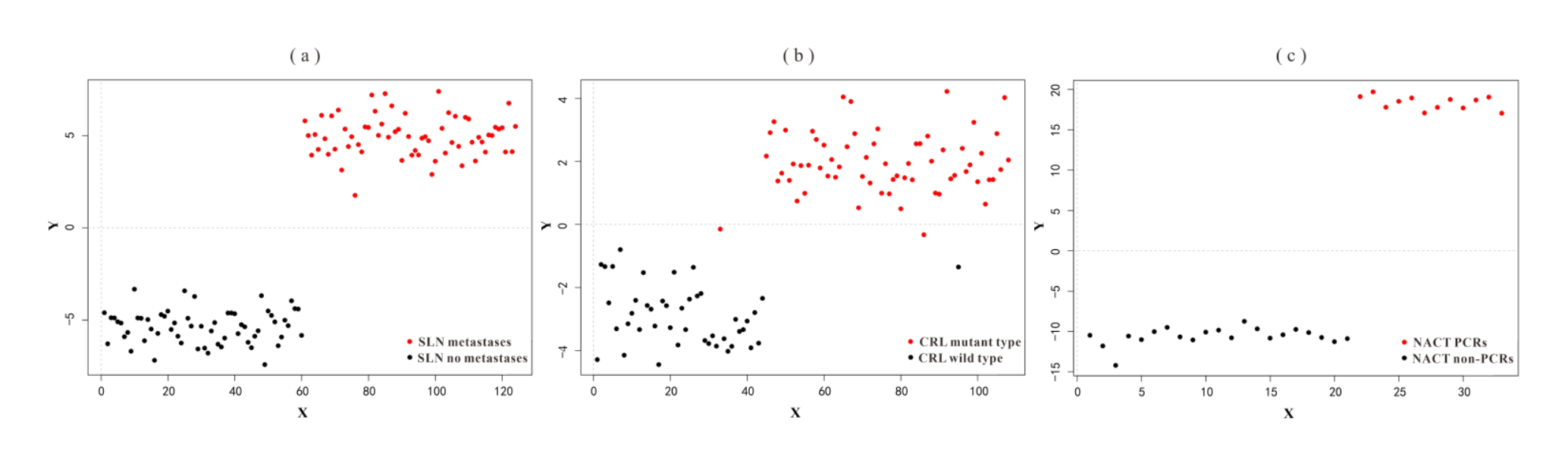}{}
\makeatother 
\caption{{(a) is the LDA visualization of data from breast cancer patients, (b) is the LDA visualization of data from patients with liver metastases from bowel cancer, and (c) is the LDA visualization of data from patients with neoadjuvant chemotherapy.}}
\label{Network}
\end{figure*}
\egroup

\bgroup
\begin{figure*}[htbp]
\centering \makeatletter\includegraphics[width=7.00in, keepaspectratio]{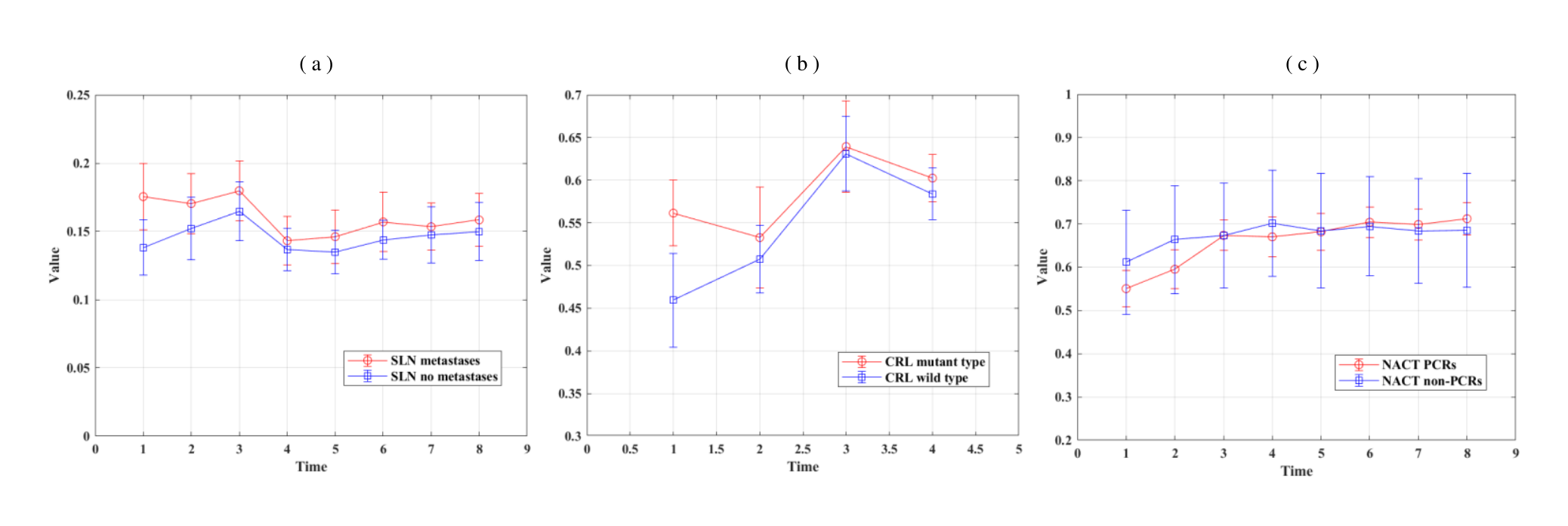}{}
\makeatother 
\caption{{(a) is the glcm\_Idn feature change in eight periods, (b) is the glszm\_SmallAreaEmphasis feature change in four periods, and (c) is the fos\_Skewness feature change in eight periods.}}
\label{Network}
\end{figure*}
\egroup

\subsection{Joint analysis and algorithm validation}

In the previous part of the experiment, we realized that a single type of algorithm feature may not be suitable for clinical prediction in more fields, so in this part of the work, we conducted a joint modeling analysis on multiple algorithm features. We used a total of three data sets with data from breast cancer patients, data from patients with liver metastases from bowel cancer, and data from a public data set to evaluate the efficacy of neoadjuvant chemotherapy. In addition, in this part of the work, we introduce a two-layer time-dependent RNN model named LSTM [44], and conduct a comparative analysis with the dynamic model designed by us. The experimental data of the RNN model are of two types: the original image and the extracted image features. The processing method of the original image is CNN (8-layer convolutional network)+RNN (2-layer LSTM), and the extracted image features are directly analyzed by using the RNN model. The accuracy results are shown in Fig. 4. Each box graph model randomly sampled the training set and test set ten times. At the same time, we used LDA to visualize the data of the three groups of dynamic models, and the visual classification effects of the models were all good. The details are shown in Fig. 5.

From the experimental results of these three kinds of data, the performance of the dynamic analysis model is still better than that of the static analysis model. In the prediction of the three types of diseases, the dynamic method has the best performance in the prediction of breast cancer, and the performance of the model is relatively stable, followed by the prediction of bowel cancer, and the performance of the model is relatively good. In the prognostic evaluation of neoadjuvant chemotherapy, the predictive effect of the model was poor, which may be due to the small amount of data. In addition, it can be seen from the comparative test of the RNN model that the performance results for breast cancer and bowel cancer are not very good. This may be because the RNN model fails to dig out the time correlation between images well and the experimental dataset used in this paper is small. The performance results in the prognostic evaluation of neoadjuvant chemotherapy are acceptable and the model is stable. For breast cancer data, we also used the CNN+RNN model to analyze and predict the original image with an accuracy rate of 0.819. The experimental results performed well, but there was still a certain gap compared with the dynamic analysis model. We also found that among these extracted features, texture features have the greatest impact on the experimental results. The specific metabolic process of the pathological tissue is mainly reflected in the texture change of the image. Moreover, through the screening of dynamic 2D and 3D features, we found that only the relative change rules of certain features at specific time points have an important impact on the specific information describing the lesion. Therefore, for these three data sets, we extracted one of the most predictive features to show how the features change over time. The details are shown in Fig. 6.

In general, by applying the dynamic features to the classification problems, we can intuitively see some changes and differences between positive and negative samples, and the dynamic features that influence different clinical problems are also quite different. For example, with respect to the prediction of sentinel lymph node metastasis in breast cancer, inverse difference normalized (IDN) [9] is one of the texture features that we extracted to assess dynamic changes. IDN normalizes the difference between the neighboring intensity values by dividing by the total number of discrete intensity values, which can denote the heterogeneity of ROIs. As shown in Fig. 6(a), in the first three periods, the gradient of patients without SLN metastasis was significantly higher than that of patients with SLN metastasis. This difference cannot be detected by static features of any given moment, but could be reflected by the RCR of discrete feature, which is one of the dynamic features proposed in this paper. This result also indicates that the primary lesion of breast cancer in SLN metastasis patients is more heterogeneous than that in patients without SLN. 

Moreover, the dynamic change trend that plays a significant role in predicting liver metastasis of colorectal cancer is small area emphasis (SAE) [11], which estimates the distribution of small regions in ROIs. A higher SAE value indicated smaller regions and fine textures. Here, the most significant difference can be observed at the slope-change trend from the first time point to the second time point(Fig. 6(b)). This result indicated that the liver lesions in CRLM patients with RAS and/or BRAF gene mutations showed a localized fine-grained structure in the precontrast phase while this difference was less obvious at the third time point, corresponding to the portal venous phase.

For the neoadjuvant chemotherapy data set, skewness, which measures the asymmetry of the distribution of values about the mean value among voxels in the ROIs [8], is one of the key static features used to evaluate changes over time. Higher skewness values indicated that there were more pixels with lower CT values; that is, the image with higher skewness values was relatively dark than that with lower skewness values. Fig. 6(c) shows that the PCR patients show a more distinct upward trend among eight time points, and the fluctuation range between different cases is small. This result indicated that patients with a steady increase in CT values at the tumor site after each treatment are likely to achieve an ideal therapeutic outcome.

\section{DISCUSSION}

The workflow of dynamic radiomics proposed in this study is an extension of traditional static radiomics in the time dimension. The extraction of all dynamic features is based on the extracted static features, which describe the change in static features at different times. This dynamic analysis includes the difference between any two time-series images; thus, most of the small changes in the lesion area would be captured by this method. Whether compared to conventional pharmacokinetics [45] or the state-of-the-art “delta-radiomics” [31], dynamic radiomics methods can provide more forms of features for building machine learning models. In fact, dynamic radiomics can be thought of as an extension and refinement based on these two types of analytical methods. For example, the peak characteristics in pharmacokinetics can be obtained indirectly by the parameter features of dynamic radiomics, which are introduced in section 3.3, while delta-radiomics describes the relative rate of change in different treatment stages, which is similar to the principle of the RCR of discrete features. However, our RCR method can obtain more features, as it describes the relative change of the images at any two time points.

\bgroup
\begin{figure}[htbp]
\centering \makeatletter\includegraphics[width=3.5in, keepaspectratio]{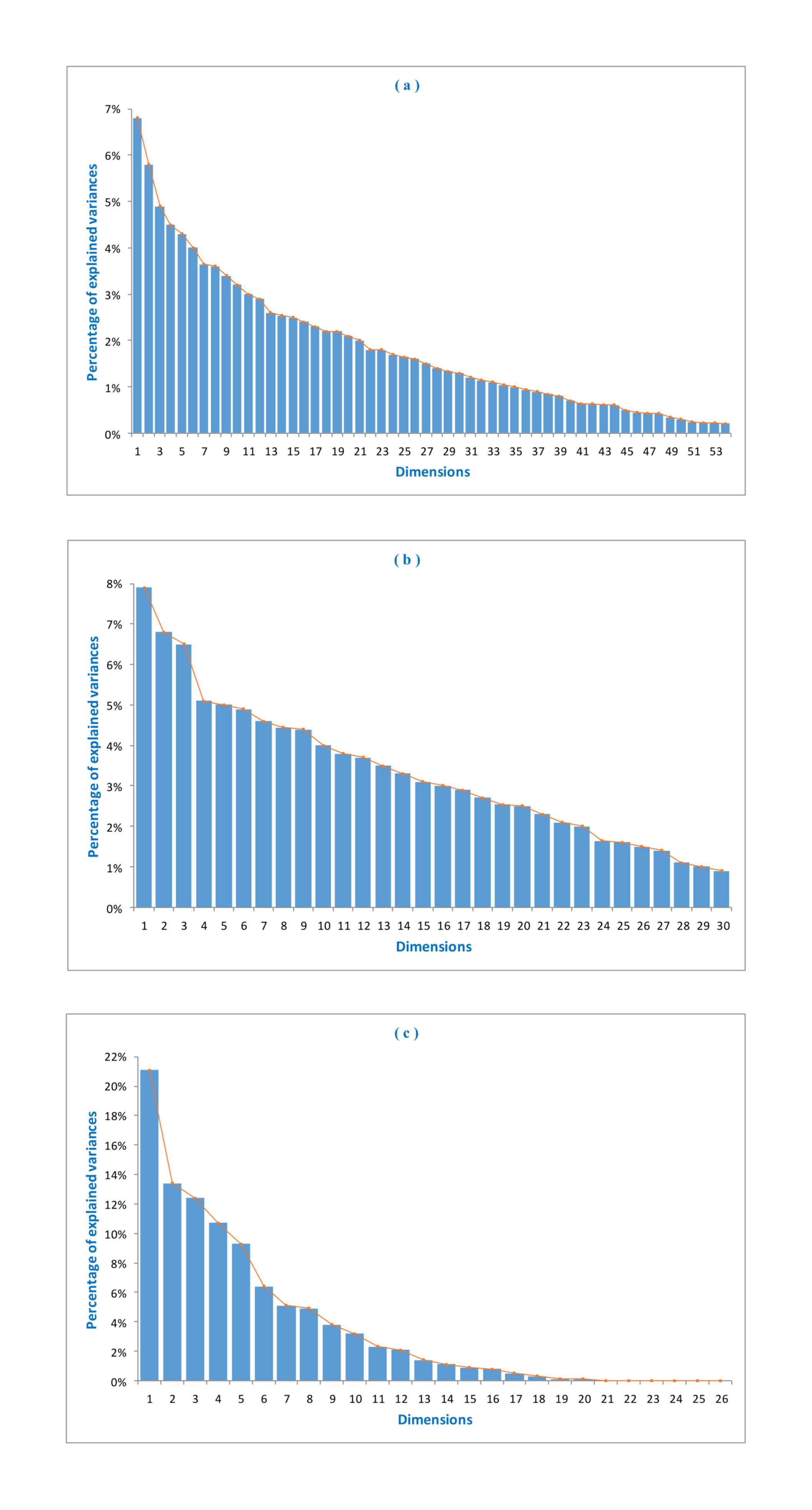}{}
\makeatother 
\caption{{(a) is the PCA scree plot of data from breast cancer patients, (b) is the PCA scree plot of data from patients with liver metastases from bowel cancer, and (c) is the PCA scree plot of data from a public data set.}}
\label{Network}
\end{figure}
\egroup

Based on the experimental results of three different clinical problems, we found that good prediction results often require more dynamic features (from 30-40) to be modeled together. Therefore, we used the PCA algorithm to continue to reduce the dimension of the dynamic features that were finally used for modeling to observe the proportion of information contained in each feature. As shown in Fig. 7, although the proportion of information contained in each principal component showed a decreasing trend after PCA, most features contributed to the classification results. As a result, the accuracy of our method is higher than of other methods because we can extract more dynamic features. 

The core contribution of this study is to propose a general framework that provides researchers in the radiomics field with a new approach to the analysis of tomographic images, which anyone can rely on to complement more dynamic feature extraction methods. In general, dynamic radiomics has the following four advantages: (1) more features: dynamic features can reflect the changes of all static features over time, so for time series medical images, this method can extract more features for model building and (2) better robustness. Dynamic features calculate the relative changes of static features; therefore, compared to static features that use absolute values, dynamic features are less likely to be affected by image quality (thickness, resolution or other imaging differences caused by different manufacturers). (3) Interpretability. Compared with deep learning features, dynamic features are interpretable, and thus more acceptable to doctors, and can be used in conjunction with various machine learning algorithms (4) Small sample set. Dynamic radiomics can use less data to obtain better results and is more suitable for small sample learning tasks.

Inevitably, there are some obvious problems with this approach. First, dynamic features are more computationally intensive than static features. If the number of static features extracted is m, and the number of dynamic features to be calculated is n, then the additional computation amount is n*m. Second, the effectiveness of feature extraction is limited by the number of time points. If there are fewer time points, it is difficult to carry out curve fitting, so the results of parameter peatures are generally not ideal. However, despite these problems, we still believe that this method has potential applications in diagnosis and prognosis assessment.

\section{Conclusion}

The dynamic radiomics proposed in this field can assist doctors in the dynamic analysis of images and provide theoretical support for clinical diagnosis. At the same time, it also provides a new way to obtain information that cannot be accomplished with static analysis.

From the perspective of clinical application, the time-varying features extracted by dynamic radiomics can indirectly reflect the specific metabolic process of the lesion tissue, thereby establishing a relative relationship between metabolic changes and pathological changes by means of computer-assisted diagnosis. This method not only expands the analysis scope of traditional radiomics but also realizes the transition from static analysis to dynamic analysis of the image and provides an effective solution for current doctors to analyze the impact of dynamic images due to subjectivity.

\section{ACKNOWLEDGMENT}

We thank Professor Liang Dong, Shenzhen Institute of Advanced Technology, Chinese Academy of Sciences, for his comments and suggestions to improve the quality of this paper.

\end{document}